\documentclass[12pt,prl]{revtex4}
\usepackage{graphicx}
\usepackage{mathrsfs}
\usepackage{amsfonts}
\usepackage{amssymb}
\usepackage{bm}

\newcommand{\beq}[2]{\begin{equation}#1\label{#2}\end{equation}}
\DeclareMathAlphabet{\mathpzc}{OT1}{pzc}{m}{it}
\newcommand{\ceq}[1]{(\ref{#1})}

\begin{document}

\title{Dynamics of two topologically entangled chains}
\author{F. Ferrari$^{1}$ J. Paturej$^{1,2}$, M. Pi\c{a}tek$^{1,3}$
 and T.A. Vilgis$^2$}
\affiliation{ $^1$ Institute of Physics, University of Szczecin,
Wielkopolska 15, 70451 Szczecin, Poland
\\$^2$ Max Planck Institute for Polymer
Research, 10 Ackermannweg,
55128 Mainz, Germany
\\
$^3$Bogoliubov Laboratory of Theoretical Physics,
Joint Institute for Nuclear Research, 141980, Dubna, Russia
}

\begin{abstract}
Starting from a given  topological invariant,
we argue that it is possible
to construct a topological field
theory with a finite number of Feynman diagrams and an amplitude of
gauge invariant objects
that is a function of that invariant.
This is for example the case of the Gauss linking number and of the
abelian BF models which has been already successfully applied in the
statistical mechanics of polymers. In this work it is shown that a
suitable generalization of the BF model can be applied also to polymer
dynamics, where the polymer trajectories are not static, but change
their shape during time.
\end{abstract}

\maketitle

\section{Introduction}

\begin{figure}[ht]
 \includegraphics[scale=0.2]{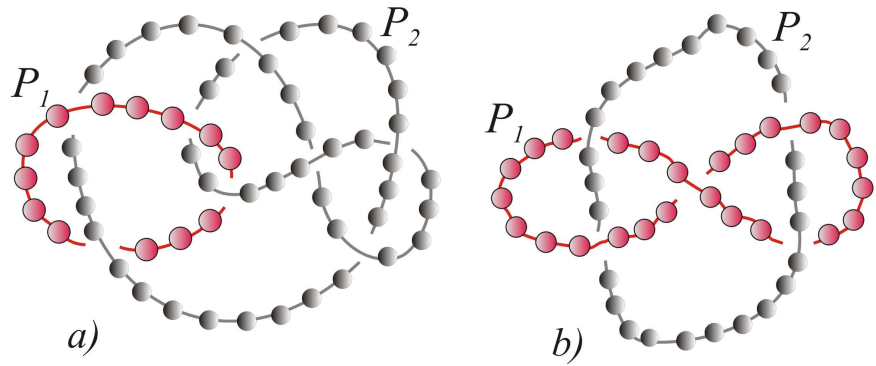}
 \caption{Entangled polymers rings $P_1$ and $P_2$ with linked trajectories
   $C_1$ and $C_2$. In
 a) polymer $P_2$ is in a nontrivial knot configuration, while in b)
both trajectories are unknots. } \label{loop}
 \end{figure}

There are many situations in which it is necessary to consider
topological relations  among one-dimensional objects that are homeomorphic
to rings. The most significant examples are provided by long flexible
polymers and biopolymers, whose trajectories may close themselves
and form what in the polymer scientific literature are called
{\it catenanes} \cite{wasser}--\cite{ff}. The latter are able to
entangle themeselves giving rise to
complex links involving two
or more interlocked chains. Additionally, each catenanae may be in the
 configuration of a nontrivial knot.
 Two cases of  polymer links
are shown in Fig.~\ref{loop}.
Besides polymers, other examples in which  topological relations
among a system of one-dimensional objects
become relevant
can be found in
  condensed matter physics (paths around defects in melted crystals)
 \cite{chaikin,pieranski}  or in particle physics (loops in quantum
 gravity and the so-called hopfions) \cite{ash,smolin,hopfions}.
In order to specify the topological states of a given system of this
kind one uses {\it knots} or {\it link invariants}. In the
following, we will be interested in the topological relations of a
system of a linked rings without taking into account the fact that
these rings could be also in a nontrivial knot configuration as for
example in Fig.~\ref{loop}~a). For this reason, we will discuss here
only link invariants.

It is well known that the correlation functions
 of the observables of a
topological field theory are topological invariants. Moreover,
the coefficients of the perturbative expansion of those
correlation functions are topological invariants too.
In practice, this means that to a finite set of Feynman diagrams
it is possible to associate a given topological invariant.
Our purpose  is  to solve the inverse problem.
This means that, starting from a given topological
invariant, we would like to obtain a topological
field theory with a finite set of Feynman diagrams and
a correlation function which is
a function of that invariant.
This is the program of topological engineering
that has been stated in Ref.~\cite{ffbookch}.
In the last few decades topological theories with the above characteristics
have been
extensively applied in the statistical mechanics of polymers, see for
instance
\cite{edwards}--\cite{ferrariTFT} and \cite{ffbookch,leal}.
The  most popular approach used
in order to distinguish the different topological
configurations of the one-dimensional objects
is based on the {\it Gauss linking number}
(GLN).
The corresponding topological field theory is an abelian BF model
discussed in Ref.~\cite{blau}.
The goal of this work is to extend this approach based
on the GLN to the case of polymer dynamics,
in which the shape of the linked trajectories is not static, but
changes in time.
\section{The topological engineering program}
The program of topological engineering in the case of links
may be summarized
as follows:\\
{\it Let $\cal{T}(\ell)$ be a link invariant, which describes the
topological properties of a $N$--component link $\ell$. It is
required that:
\begin{itemize}
\item[a)] the invariant $\cal{T}(\ell)$ is explicitly written
as a functional of trajectories $C_1,\ldots,C_N$ of knots
composing the link.\\
Given a link invariant of this kind, find a topological
field theory with observables ${\cal{O}}_1,\ldots,{\cal{O}}_n$
such that $\cal{T}(\ell)$, or equivalently a function
$F[\cal{T}(\ell)]$ of it, can be expressed as the correlator
of these observables
\beq{
F({\cal{T}}) = \int\!{\cal{D}}\lbrace \phi \rbrace e^{-S(\lbrace \phi \rbrace )}
{\cal{O}}_1(\lbrace \phi\rbrace),\ldots,{\cal{O}}_n(\lbrace \phi\rbrace)
}{corr}
where $S(\lbrace\phi \rbrace)$ is the action of a system and $\lbrace \phi\rbrace $
is a set of fields that can be scalars, vectors or higher order tensors.
\end{itemize}
The topological field theory and its observables should satisfy the following
conditions:
\begin{itemize}
\item[b)] Each observable ${\cal{O}}_i$, $i=1,\ldots,n$,
must depend on the trajectory of only one knot
\item[c)] No further regularization should be necessary in order
to compute the correlator $\langle{\cal O}_1,\ldots,{\cal O}_n\rangle$,
apart from the usual regularization schemes required by the possible
presence of ultraviolet divergences.
\end{itemize} }

An example of topological engineering is based on the GLN and the
abelian BF field theory. The GLN is given by:
\beq{
\chi(C_1,C_2) = \frac 1{4\pi}\epsilon_{\mu\nu\rho}
\oint_{C_1}\!dx_1^{\mu}(s_1) \oint_{C_2}\!dx_2^{\nu}(s_2)
\frac {(x_1(s_1) - x_2(s_2) )^{\rho} }{|x_1(s_1) - x_2(s_2)|^3}
}{gln}
where $x_1(s_1)^{\mu}$ and $x_2(s_2)^{\nu}$ are spatial curves in
three dimensions
that represent respectively the closed trajectories $C_1$ and $C_2$ of
two polymers $P_1$ and $P_2$.
The Greek indexes $\mu,\nu,\rho=1,2,3$ denote the spatial components.
Here $s_1$ and $s_2$ represent the arc-lengths on the curves $C_1$ and $C_2$.
$s_1$ and $s_2$ are defined in a such a way that
$0\leq s_1\leq L$ and  $0\leq s_2\leq L$.
To find a field theory which is associated to the invariant
$\chi(C_1,C_2)$, we rewrite \ceq{gln}
as follows
\beq{
\chi(C_1,C_2) = \int\!d^3x \int\!d^3y
\xi_1^{\mu}(x)G_{\mu\nu}(x-y)\xi_2^{\nu}(y)
}{gln_alt}
where
\beq{
\xi_1^{\mu}(x) = \oint_{C_1}\!dx_1^{\mu}\delta(x-x_1) \qquad
\xi_2^{\nu}(x) = \kappa\oint_{C_2}\!dx_2^{\nu}\delta(x-x_2)}
{currents}
are called the  bond vectors densities and
\beq{G_{\mu\nu}(x-y) = \frac 1{2\pi\kappa}\epsilon_{\mu\nu\rho}
\frac{(x-y)^{\rho}}{|x-y|^3}
}{propagator}
Let us note that
$G_{\mu\nu}(x-y)$ coincides with the propagator of the abelian BF
model discussed in Ref.~\cite{blau}. To make the connection with the
BF model even more explicit,
we have introduced a new parameter $\kappa$, which
will play later the role of the coupling constant of that model.
Clearly, the addition of this parameter is irrelevant. As a matter of
fact, the right hand side of Eq.~\ceq{gln_alt} does not depend
on $\kappa$.
Now the quantity
\beq{
e^{i\chi(C_1,C_2)} = e^{i\int\!d^3x  \int\!d^3y \xi_1^{\mu}(x)G_{\mu\nu}(x-y)\xi_2^{\nu}(y)}
}{genfun}
can be regarded as the generating functional of a Gaussian field
theory with
propagator $G_{\mu\nu}(x-y)$ for the very special choice of currents
\ceq{currents}.
It is easy to recognize that the underlaying
field theory is an Abelian BF model
with action
\beq{
S_{\mbox{\tiny BF}} = i\kappa \epsilon^{\mu\nu\rho} \int\!d^3x  A_{\mu}\partial_{\nu}B_{\rho}
}{BFaction}
It is possible to show that the abelian version of the BF model is
actually equivalent to two
Abelian C-S field theories.
If we quantize the above topological field theory using the Lorentz
gauge fixing,
in which both fields
$A_{\mu}$ and $B_{\mu}$ are completely transverse,
we obtain the
following relation
\beq{
e^{i\chi(C_1,C_2)} =
\int\!{\cal D}A_{\mu}{\cal D}B_{\mu} e^{-S_{\mbox{\tiny BF}}}
e^{i\int\!d^3x \xi_1^{\mu} A_{\mu}}e^{i\kappa\int\! d^3x \xi_2^{\mu}B_{\mu}}
\delta(\partial^{\mu}A_{\mu})\delta(\partial^{\mu}B_{\mu})
}{genfunCS}
The above equation is the analog of Eq.~\ceq{corr} in the present case. There are
just two observables ${\cal O}_1$ and ${\cal O}_2$, namely the
 two Abelian Wilson loops given below:
\beq{
{\cal O}_1 = e^{i\int\!d^3x \xi_1^{\mu} A_{\mu} } \qquad
{\cal O}_2 = e^{i\kappa\int\!d^3x \xi_2^{\mu} B_{\mu} }
}{wilson}

\section{The case of dynamics}

In this Section we would like to extend the program of topological
engineering
 to the case of two trajectories
whose configurations are changing during time. This problem is very
important
to study the dynamics of two entangled polymers.
Once again, we choose the Gauss linking invariant in order to impose
topological conditions on two closed trajectories $C_1$ and $C_2$.
The only difference from the previous static example is that now the
curves $x_1$ and $x_2$  depend on time, i.e. $x_1=x_1(t,s_1)$
and $x_2=x_2(t,s_2)$.
The GLN can still be defined, but will be a time dependent quantities:
\beq{
\chi(t,C_1,C_2) = \frac 1{4\pi} \epsilon_{\mu\nu\rho}
\oint_{C_1}\!dx_1^{\mu}(t,s_1) \oint_{C_2}\!dx_2^{\nu}(t,s_2)
\frac {(x_1(t,s_1) - x_2(t,s_2) )^{\rho} }{|x_1(t,s_1) - x_2(t,s_2)|^3}
}{chitime}
Of course, if the trajectories would be impenetrable, then $\chi$
would be a
constant,  since it is not possible to change the topological
configuration of a system
of knots if their trajectories are not allowed to cross
themselves. However, in
the absence of excluded volume interactions
models of polymer physics are phantom, i.e. crossings are allowed.
For  this reason, we will require  that only the
time average of the GLN
is fixed. As a consequence, we will consider a time  averaged version
 of the GLN on the
time interval
$[0,t_f]$:
\beq{
\langle \chi (t,C_1,C_2) \rangle =
\int_0^{t_f} \frac {dt}{t_f} \chi (t,C_1,C_2)
}{timeaGLN}

Next, we generalize Eq.~\ceq{genfun} to the case of dynamics. To this
purpose, we introduce the following field theory
\beq{
S = \frac 1{t_f} \epsilon_{\mu\nu\rho} \int\!d\eta d^3x
 A^{\mu}(\eta,x)\partial_x^{\nu}B^{\rho}(\eta,x)
}{actiondyn}
The above action differs from that of Eq.~\ceq{BFaction} by the addition
of the fourth dimension represented by variable $\eta$, with
$-\infty<\eta<+\infty$. Note that $S$ is not invariant under
diffeomorphism on the whole dimensional space spanned by the
coordinates $x^1,x^2,x^3$ and $\eta$, but only on its
three dimensional spatial section.
As a consequence, strictly speaking $S$ does not describe a
topological field theory.
The propagator corresponding to the action \ceq{actiondyn}
in the Lorentz gauge is given by
\beq{
G_{\mu\nu}(\eta,\eta';x,x') = \frac {t_f}{2\pi} \epsilon_{\mu\nu\rho}
\frac{(x-x')^{\rho}}{|x-x'|^3} \delta(\eta - \eta')
}{propdyn}
The analog of Eq.~\ceq{genfun} is
\beq{
e^{-i\lambda\chi(C_1,C_2)} = \int\!{\cal{D}} A_{\mu}{\cal{D}}B_{\nu} e^{-iS}
e^{-i\int\! d\eta d^{3}x  ( J_1^{\mu}(\eta,x)A_{\mu}(\eta,x) + J_2^{\mu}(\eta,x)B_{\mu}(\eta,x)  ) }
}
{genfungen}
where
\beq{J_1^{\mu}(\eta,x) = \frac{1}{2t_f} \int_0^{t_f}\!\frac{dt}{t_f}\delta(\eta - t)
\int_0^{L_1}\!ds_1 \frac{\partial }{\partial s_1}x_1^{\mu}(t_1,s_1)
\delta^{(3)}(x-x_1(t,s_1))
}{prad1}
and
\beq{
J_2^{\mu}(\eta,x) = \lambda \int_0^{t_f}\!\frac{dt}{t_f}\delta(\eta - t)
\int_0^{L_2}\!ds_2 \frac{\partial }{\partial s_2}x_2^{\mu}(t_1,s_2)
\delta^{(3)}(x-x_2(t,s_2))
}{prad2}
The right hand side of Eq.~\ceq{genfungen} can be seen as the amplitude of the two observables
\beq{
{\cal O}_1 = e^{-i\int\! d\eta d^3x J_1^{\mu}(\eta,x)A_{\mu}(\eta,x)} \qquad
{\cal O}_2 = e^{-i\int\! d\eta d^3x J_2^{\mu}(\eta,x)B_{\mu}(\eta,x)}
}{canbeseen}
To prove Eq.~\ceq{genfungen} it is sufficient to perform the Gaussian
integration in the fields $A^{\mu}$ and $B^{\mu}$. The result of that
operation is
\beq{
e^{-i\int\! d\eta d^{3}x  ( J_1^{\mu}(\eta,x)A_{\mu}(\eta,x) + J_2^{\mu}(\eta,x)B_{\mu}(\eta,x)  ) }=
e^{-i\int\!d\eta d^3x\int\!d\eta' d^3x'  J_1^{\mu}(\eta,x) G_{\mu\nu}(\eta,\eta';x,x') J_2^{\nu}(\eta',x') }
}{gaussintegration}
Using the explicit expression of the propagator $G_{\mu\nu}(\eta,\eta';x,x')$
given in Eq.~\ceq{propdyn} it is possible to verify Eq.~\ceq{genfungen}
after eliminating the spurious variables $\eta,\eta'$ and $x,x'$:
\begin{eqnarray}
&&e^{-i\int\!d\eta d^3x\int\!d\eta' d^3x'  J_1^{\mu}(\eta,x) G_{\mu\nu}(\eta,\eta';x,x') J_2^{\nu}(\eta',x') } =\\
\nonumber
&&\!\!\!\!\!\!\exp{\left[-\frac {i\lambda}{4\pi}\int_0^{t_f}\!\frac{dt}{t_f} \int_0^{L_1}\!ds_1\int_0^{L_2}\!ds_2
\epsilon_{\mu\nu\rho}
\frac{\partial}{\partial s_1}x_1^{\mu}(t,s_1) \frac{\partial}{\partial s_2}x_2^{\nu}(t,s_2)
\frac{(x_1(t,s_1) - x_2(t,s_2))^{\rho}}{|(x_1(t,s_1) - x_2(t,s_2)|^3}
\right]}
\label{endofstory}
\end{eqnarray}
The right hand side of above equation coincides with
$e^{-i\lambda\chi(C_1,C_2)}$. This
completes our proof.

\section{Concluding remarks}

In this work the program of topological engineering has been extended to the
case of the dynamics of two polymer chains. In particular, the Gauss linking
invariant has been considered. It has been shown that a time average version
of this topological invariant can be reproduced from an amplitude of
a field theory in the form of Eq.~\ceq{corr}. This amplitude is given in
Eq.~\ceq{genfungen}. Due to the fact that  the conformations of the chains
change during time, the underlying field theory is four dimensional
and it is topological only with respect to diffeomorphisms of
the spatial section of four dimensional space.

\section{Acknowledgments}
One of us -- M. Pi\c{a}tek -- would like to thank the University of
Szczecin and the Faculty of Mathematics and Physics of that
University for the kind hospitality.

\end{document}